# Convolutional Compressed Sensing Using Deterministic Sequences

Kezhi Li, *Student Member, IEEE,* Lu Gan, *Member, IEEE,* and Cong Ling, *Member, IEEE*

*Abstract*—In this paper, a new class of circulant matrices built from deterministic sequences is proposed for convolution-based compressed sensing (CS). In contrast to random convolution, the coefficients of the underlying filter are given by the discrete Fourier transform of a deterministic sequence with good autocorrelation. Both uniform recovery and non-uniform recovery of sparse signals are investigated, based on the coherence parameter of the proposed sensing matrices. Many examples of the sequences are investigated, particularly the Frank-Zadoff-Chu (FZC) sequence, the *m*-sequence and the Golay sequence. A salient feature of the proposed sensing matrices is that they can not only handle sparse signals in the time domain, but also those in the frequency and/or or discrete-cosine transform (DCT) domain.

## I. Introduction

COMPRESSED sensing (CS) is a growing theory in signal processing aiming at efficient sampling of signals [1], [2]. Consider a length-$N$ signal $\mathbf{x}$ and suppose that the basis $\mathbf{\Psi}$ provides a $K$-sparse representation of $\mathbf{x}$. That is, $\mathbf{x} = \mathbf{\Psi}\mathbf{f}$, where $\mathbf{f}$ can be approximated using only $K \ll N$ non-zero entries and $\mathbf{\Psi}$ is referred to as the sparsifying transform. Throughout this paper, we assume that $\mathbf{\Psi}$ is an $N \times N$ normalized unitary matrix satisfying $\mathbf{\Psi}^*\mathbf{\Psi} = \mathbf{I}_N$. The data acquisition process in CS can be described as

$$\mathbf{y} = \mathbf{\Phi}\mathbf{x} + \mathbf{e} = \mathbf{\Phi}\mathbf{\Psi}\mathbf{f} + \mathbf{e}, \qquad (1)$$

where $\mathbf{y}$ represents an $M \times 1$ sampled vector, $\mathbf{\Phi}$ is an $M \times N$ measurement/sensing matrix and $\mathbf{e}$ is a noise vector. It was shown in [1], [2] that if $\mathbf{\Phi}$ is a Gaussian or Bernoulli random operator, $\mathbf{x}$ can be faithfully recovered from $\mathbf{y}$ using nonlinear optimization provided that $M \geq \mathcal{O}(K\log(N/K))$.

Although Gaussian or Bernoulli operators offer optimal theoretical bounds, they require huge memory for storage and high computational cost for signal reconstruction. Besides, fully random matrices are often difficult or expensive in hardware implementation. Taking these issues into account, many researchers have investigated structured CS operators [3]–[10]. Among them, one class of fast CS sampling operators is realized by convolving the signal of interest with a random filter and then subsampling [4], [5], [8], [9], [11]. They hold great potential in applications such as sparse channel estimation [4], Fourier optics [8], Radar imaging [8], [11] and coded aperture imaging [12].

Note that for convolution-based CS, most existing works focus on filters with independent and identically distributed (i.i.d.) random coefficients. In this paper, we propose a new framework by convolution with a *deterministic* filter followed by random sampling of the outputs. The filter is constructed from a sequence with good autocorrelation property, such that its energy is spread out in the time domain and uniformly distributed across the (discrete) spectrum. A deterministic filter is more convenient to implement than a random one. Besides, deterministic sequences with good correlation properties have found wide applications in Radar, communications and imaging. By demonstrating that these sequences can be used under the convolutional CS framework, it may lead to more efficient hardware/software design in above mentioned applications. In particular, we show that the proposed scheme can efficiently sample a sparse signal in the time/spatial ($\mathbf{\Psi} = \mathbf{I}_N$) or spectral ($\mathbf{\Psi} = \frac{1}{\sqrt{N}}\text{IFFT}$) domain. Specifically, for *all* $K$-sparse signals of length $N$ in the time or spectral domain, robust reconstruction can be achieved when the number of measurements satisfies $M \geq \mathcal{O}\left(K\log^4 N\right)$, while for *any* given $K$-sparse signal, it can be recovered from only $M \geq \mathcal{O}(K\log N)$ measurements. In addition, when the filter is constructed from the Frank-Zadoff-Chu (FZC) sequence, these results also hold for sparse signals in the discrete cosine transform (DCT) domain. In many applications, it is highly desirable that a data acquisition system can work efficiently for both time/spatially sparse signals and spectrally sparse ones. For example, in military communications, both time-hopping (time-domain sparse) and frequency-hopping signals (spectrally sparse) are used to achieve the capability of anti-jamming or low probability of intercept. As another example, in Radar imaging [13], a set of $K$ point targets are spatially sparse, while a smooth target is sparse in the Fourier domain. Moreover, in astronomical imaging [14], an image could be spatially sparse (with only a few bright objects in a dark background) or spectrally sparse as most natural images. Last but not the least, in compressive video imaging, each frame can be sparsified by the DCT transform. Due to strong temporal correlations, the residue images (after motion compensation) are spatially sparse.

The rest of this paper is organized as follows. Section II

The material in this paper was partially presented at IEEE Information Theory Workshop (ITW'09), Oct. 2009, Taormina, Italy, and at International Conference on Acoustics, Speech and Signal Processing (ICASSP'11), May 2011, Prague, Czech. This work was supported in part by the UK MOD University Defence Research Centre (UDRC) in Signal Processing and by the UK EPSRC under Grant EP/I038853/1.

K. Li and C. Ling are with the Department of Electrical and Electronic Engineering, Imperial College London, London, SW7 2AZ, UK (e-mail: k.li08@imperial.ac.uk, cling@ieee.org).

L. Gan is with the School of Engineering and Design, Brunel University, London UB8 3PH, UK (e-mail: lu.gan@brunel.ac.uk).

gives a brief review of CS theory and in particular, random convolution-based CS. Section III introduces the framework of the proposed system, its potential applications and summarizes the main results of this paper. Sections IV is devoted to the coherence analysis of complex and real-coefficient filters built from three classes of deterministic sequences, namely, the polyphase sequences, maximum-length sequences and Golay complementary sequences. Simulation results are given in V, followed by conclusions in Section VI.

*Notation:* In this paper, bold letters are used to denote a vector or a matrix. For an $M \times N$ matrix $\mathbf{A}$, $\mathbf{A}(p,q)$ ($0 \leq p \leq M-1$, $0 \leq q \leq N-1$) represents the element on its $p$-th row and $q$-th column. $\mathbf{A}^T$ and $\mathbf{A}^*$ denote the transpose and Hermitian transpose of $\mathbf{A}$, respectively. For an $N \times N$ given matrix $\mathbf{A}$, we denote by $\mu(\mathbf{A})$ its coherence parameter, i.e., the maximum magnitude,

$$\mu(\mathbf{A}) = \max_{0 \leq p,q \leq N-1} |\mathbf{A}(p,q)|.$$

For two $N \times N$ matrices $\mathbf{A}$ and $\mathbf{B}$, their mutual coherence $\mu(\mathbf{A}, \mathbf{B})$ is defined as

$$\mu(\mathbf{A}, \mathbf{B}) = \mu(\mathbf{AB}) = \max_{0 \leq p,q \leq N-1} |\mathbf{A}(p,:)\mathbf{B}(:,q)|,$$

where $\mathbf{A}(p,:)$ and $\mathbf{B}(:,q)$ correspond to the $p$-th row of $\mathbf{A}$ and $q$-th column of $\mathbf{B}$, respectively. $\mathbf{F}$ is the $N \times N$ fast Fourier transform (FFT) matrix where $\mathbf{F}(p,q) = e^{-\frac{2\pi j pq}{N}}$ ($0 \leq p, q \leq N-1$). We use the standard asymptotic notation $f(x) = \mathcal{O}(g(x))$ when $\limsup_{x \to \infty} |f(x)/g(x)| < \infty$.

## II. REVIEW OF COMPRESSED SENSING

In this section, we first review uniform and non-uniform recovery in compressed sensing, and in particular theoretical performance bounds for randomly subsampled unitary matrices. We then highlight existing works of random convolution-based CS.

### A. Uniform vs. Non-Uniform Recovery

Let $\mathbf{\Theta} = \mathbf{\Phi}\mathbf{\Psi}$. Then (1) can be re-written as

$$\mathbf{y} = \mathbf{\Theta}\mathbf{f} + \mathbf{e}. \tag{2}$$

Hence, reconstruction of $\mathbf{x} = \mathbf{\Psi}\mathbf{f}$ is equivalent to recovery of a $K$-sparse vector $\mathbf{f}$. Note that as $M < N$, eq. (2) is in general under-determined. To recover $\mathbf{f}$ (or equivalently, $\mathbf{x}$) from $\mathbf{y}$, nonlinear optimization is required. In the noiseless case (i.e, $\mathbf{e} = \mathbf{0}$), exact recovery can be achieved by a standard $l_1$ minimization program [15]

$$\min \|\mathbf{f}\|_{l_1} \quad \text{s.t.} \quad \mathbf{y} = \mathbf{\Theta}\mathbf{f}. \tag{3}$$

In the noisy case, $\mathbf{f}$ can be reconstructed using the unconstrained LASSO [16] that solves the $l_1$ regularized square problem

$$\min \lambda \|\mathbf{f}\|_{l_1} + \frac{1}{2} \|\mathbf{y} - \mathbf{\Theta}\mathbf{f}\|^2, \tag{4}$$

where $\lambda$ is the Lagrangian constant. In addition to $l_1$-based algorithms, many greedy algorithms, such as orthogonal matching pursuit [17], subspace pursuit [18], CoSaMP [19] and their variants have been proposed for sparse signal reconstruction.

These algorithms require lower computational complexity than $l_1$-based optimization with somewhat weaker theoretical guarantees.

In CS reconstruction, *uniform recovery* [2] means that once the sampling operator $\mathbf{\Phi}$ is constructed, all sparse signals in a certain basis $\mathbf{\Psi}$ can be recovered as long as $M$ is sufficiently large. To achieve uniform recovery, many recovery algorithms require the restricted isometry property (RIP).

*Definition 1 (RIP):* An $M \times N$ matrix $\mathbf{\Theta} = \mathbf{\Phi}\mathbf{\Psi}$ is said to satisfy the RIP with parameters $(K, \delta)$ ($\delta \in (0, 1)$) if

$$(1-\delta)\|\mathbf{f}\|^2 \leq \|\mathbf{\Theta}\mathbf{f}\|^2 \leq (1+\delta)\|\mathbf{f}\|^2, \text{ for all } \mathbf{f} \in \Gamma, \tag{5}$$

where $\Gamma$ represents the set of all $K$-sparse vectors of length $N$.

For RIP constant $\delta$ required in different sparse recovery algorithms, please refer to [7] for details. Note that the RIP is a very strong restriction. Among existing sampling operators, it is known that the i.i.d. Gaussian and Bernoulli matrices satisfy the RIP when $M \geq \mathcal{O}(\delta^{-2} K \log N)$. However, as we have mentioned before, these full random operators are impractical for large-scale CS data acquisition. Another subclass of operators satisfying the RIP is randomly sampled unitary matrix, as summarized in the following theorem [7], [20].

*Theorem 1 (RIP for randomly subsampled unitary matrix):* Suppose that the $M \times N$ matrix $\mathbf{\Theta}$ is a randomly subsampled unitary matrix, i.e., it can be written as $\mathbf{\Theta} = \frac{1}{\sqrt{M}} \mathbf{R}_\Omega \mathbf{U}$, where $\frac{1}{\sqrt{M}}$ is a normalizing coefficient, $\mathbf{R}_\Omega$ is a random sampling operator which selects $M$ samples out of $N$ ones uniformly at random, and $\mathbf{U}$ is an $N \times N$ unitary matrix satisfying $\mathbf{U}^* \mathbf{U} = N \mathbf{I}_N$. Then $\mathbf{\Theta}$ satisfies the RIP with high probability provided that [7], [20]

$$M \geq \mathcal{O}\left(\delta^{-2} \mu^2(\mathbf{U}) K \log^4 N\right). \tag{6}$$

The above theorem implies that the RIP bound of a randomly subsampled unitary matrix depends on $\mu(\mathbf{U})$. Note that for a unitary matrix $\mathbf{U}$ with $\mathbf{U}^*\mathbf{U} = N\mathbf{I}_N$, we have $1 \leq \mu(\mathbf{U}) \leq \sqrt{N}$. In case when $\mathbf{U}$ is chosen as the FFT or the Walsh-Hadamard transform (WHT), $\mu(\mathbf{U}) = 1$ and by Eq. (6), we have

$$M \geq \mathcal{O}\left(\delta^{-2} K \log^4 N\right). \tag{7}$$

One can also observe that compared with the optimal bound provided by fully random matrices, there is an extra $\log^3 N$ factor in (7). To address this issue, several researchers have relaxed the conditions and investigated the case of non-uniform recovery, where one just needs to reconstruct a given sparse signal. Theorem 2 presents the results for non-uniform recovery of a randomly subsmapled unitary matrix using $l_1$-based optimization.

*Theorem 2 (Non-uniform recovery):* Assume that $\mathbf{\Theta}$ is a randomly subsampled unitary matrix that follows the same definition as in Theorem 1. Let $\mathbf{f}$ in (2) be a fixed arbitrary $K$-sparse signal. Then $\mathbf{f}$ can be faithfully recovered from $\mathbf{y}$ using $l_1$-based optimization (i.e., (3) in the noiseless case and (4) in the noisy case) if $M$ satisfies [21]

$$M \geq \mathcal{O}(\mu^2(\mathbf{U}) K \log N). \tag{8}$$

Theorem 2 implies that using randomly subsampled FFT or WHT, the number of samples required for non-uniform reconstruction is nearly optimal. This is because that non-uniform recovery is much weaker than uniform recovery. It should also be noted that the above non-uniform recovery only holds for $l_1$ optimization. It is still unknown whether we can get similar bounds for fast greedy recovery algorithms such as subspace pursuit [18] and CoSaMP [19].

At this point, it is worth pointing out that although partial FFT (or WHT) has near-optimal theoretical guarantee, easy hardware implementation and fast-computable recovery, its major shortcoming is the lack of *universality*. A universal sensing matrix means that it can handle signals that are sparse on any $\Psi$. If $\Phi$ is a Gaussian random matrix, the matrix $\Phi\Psi$ will remain Gaussian for any unitary transform $\Psi$. However, if $\Phi$ is randomly sampled from a FFT, it will not be universal, as $\mu(\mathbf{F}\Psi)$ can not be $\mathcal{O}(1)$ for all bases $\Psi$. As a remedy, in this paper, we will propose a new class of randomly subsampled circulant matrices that can be used to efficiently sample sparse signals in either the time or frequency domain.

### B. Random Convolution for CS

Tropp *et al.* [5], [22] first proposed the idea of CS using convolution with an i.i.d. sequence followed by fixed regular sampling. The effectiveness of such an approach has been demonstrated through numerical simulations. Later, many people have investigated cyclic convolution with an $N$-tap random filter [4], [8], [9], in which the sampling operator $\Phi$ can be represented as a partial circulant matrix with the following form

$$\Phi = \frac{1}{\sqrt{M}}\mathbf{R}_\Omega \mathbf{A} \qquad (9)$$

where $\mathbf{A}$ is a circulant matrix that can be expressed as

$$\mathbf{A} = \begin{bmatrix} a_0 & a_{N-1} & \cdots & a_1 \\ a_1 & a_0 & \cdots & a_2 \\ \vdots & \vdots & \ddots & \vdots \\ a_{N-1} & a_{N-2} & \cdots & a_0 \end{bmatrix}. \qquad (10)$$

For $\Phi$ given in (9), the measurement process can be realized by circularly convolving $\mathbf{x}$ with a filter $\mathbf{a} = \begin{bmatrix} a_0 & a_1 & \cdots & a_{N-1} \end{bmatrix}^T$ and then downsample the output at locations indexed by $\Omega$. It is also well known that a circulant matrix $\mathbf{A}$ can be diagonalized using FFT as follows

$$\mathbf{A} = \frac{1}{\sqrt{N}}\mathbf{F}^*\Sigma\mathbf{F}, \qquad (11)$$

where $\Sigma = \text{diag}(\sigma) = \text{diag}(\sigma_0, \sigma_1, \cdots, \sigma_{N-1})$. Eq. (11) suggests that a circulant operator is fast computable. It is easy to see that the filter vector $\mathbf{a}$ (i.e., the first column of $\mathbf{A}$) can be obtained by taking the IFFT[1] of the diagonal sequence $\sigma = \begin{bmatrix} \sigma_0 & \sigma_1 & \cdots & \sigma_{N-1} \end{bmatrix}^T$, i.e.,

$$\mathbf{a} = \frac{1}{\sqrt{N}}\mathbf{F}^*\sigma. \qquad (12)$$

[1] For convenience, the definition here differs from the standard one IFFT $= \frac{1}{N}\mathbf{F}^*$ by a factor of $1/\sqrt{N}$.

In other words, $\sigma$ is the normalized FFT of $\mathbf{a}$. It is clear that when $\sigma$ is a unimodular sequence, i.e., $|\sigma_k| = 1$ ($0 \leq k \leq N-1$), $\mathbf{A}$ is a unitary matrix satisfying $\mathbf{A}^*\mathbf{A} = N\mathbf{I}_N$.

In existing works, the coefficient vector $\mathbf{a}$ is constructed randomly. In [4], [9], $\mathbf{a}$ is a binary random sequence, where each $a_i$ takes the values of $+1$ and $-1$ with equal probability. An alternative way is to obtain $\mathbf{a}$ from $\sigma$, which is a binary random sequence [9] or a unimodular sequence with random phases [8], i.e., $\sigma_k = e^{j\theta_k}$, where $\theta_k$ is a random variable that is uniformly distributed in $[0, 2\pi)$.

The sampling operator $\mathbf{R}_\Omega$ can be either deterministic or random, as summarized below.

*1) Deterministic subsampling:* In deterministic sampling, $\Omega$ is chosen as any arbitrary subset of $\{1, 2, \cdots, N\}$ with cardinality $|\Omega| = M$. It was shown in [9] that $\Phi$ given by (9) satisfies RIP with parameters $(K, \delta)$ provided that $M \geq \mathcal{O}\left((K\log N)^{\frac{3}{2}}\right)$. More recently, this bound has been improved to $M \geq \mathcal{O}\left((K\log^4 N)\right)$ [23]. Non-uniform recovery results have been investigated in [24], where the author considered the recovery of a given $K$-sparse signal whose nonzero components have random signs. Under this model, it was established in [23] that exact recovery can be achieved using $l_1$ optimization when $M \geq \mathcal{O}(K\log N)$. However, unlike Theorem 2, this bound only holds for noiseless measurement and hence the guarantee for stable recovery under noisy measurements is unclear. More importantly, as we will show later, when $\mathbf{R}_\Omega$ is a deterministic operator, $\Phi$ given in (9) works poorly for a spectrally sparse signal, which implies that it cannot be used directly to sample a natural image (which is often sparse in the DCT or the wavelet domain).

*2) Random subsampling:* To achieve a universal convolution-based CS, Romberg [8] proposed to use a random sampling operator $\mathbf{R}_\Omega$. Note that if $\sigma$ is a random unimodular sequence, the coherence parameter of $\mathbf{A}$ given by (11) satisfies

$$\mu(\mathbf{A}\Psi) = \mathcal{O}(\sqrt{\log N}). \qquad (13)$$

By Theorem 1, such a universal operator satisfies the RIP when $M \geq \mathcal{O}(\delta^{-2}K\log^5 N)$ and by Theorem 2, $M \geq \mathcal{O}(\delta^{-2}K\log^2 N)$ measurements are required for non-uniform recovery. Note that compared with the optimal bounds offered by a randomly subsampled unitary matrix, there is an extra $\log N$ factor in random convolution. It is thus natural to ponder: Can we get better bounds for convolution-based CS systems with random sampling?

## III. DETERMINISTIC FILTER FOR CONVOLUTIONAL CS

In this Section, we propose a new framework which answers the afore-posed question in the affirmative.

### A. Problem Formulation

Unlike previous work, we propose the use of a *deterministic filter* followed by random sampling for convolution-based CS. Specifically, for $\Phi$ given in (9), $\mathbf{R}_\Omega$ is a random sampling operator and $\mathbf{A}$ is a deterministic circulant matrix. Just as in a random filter, there are two ways to construct a deterministic

**A**. In the *frequency-domain* approach, $\sigma$ is constructed first and **a** is obtained from (12). It is clear that **A** is a unitary matrix if and only if $\sigma$ is a unimodular sequence. In the *time-domain* approach, the filter vector **a** is constructed directly. Since it is not easy to get a unitary **A** using the time-domain approach, we will mainly focus on the frequency-domain approach, in which $\sigma$ is a unimodular sequence. An example of the time-domain construction will be given in Section IV-B. Note that when $\sigma$ is a unimodular sequence, **A** in general is complex valued. Yet real-valued filters are desired in many applications, such as Fourier optics and coded aperture imaging [12]. To generate real sensing matrices, $\sigma$ needs to satisfy the following conjugate symmetry condition [8]:

$$\sigma_k = \sigma_{N-k}^*, \quad 1 \leq k \leq N-1, \tag{14}$$

where the superscript $*$ represents the complex conjugate operation. The following theorem presents the requirements of a deterministic $\sigma$ under the CS framework:

*Theorem 3:* Consider a CS sampling operator $\mathbf{\Phi}$ given in (9), where $\mathbf{R}_\Omega$ is a random sampling operator and the unitary circulant matrix **A** is generated from (11), with $\sigma = [\sigma_0, \sigma_1, \cdots, \sigma_{N-1}]^T$, $|\sigma_k| = 1$, $k = 0, 1 \cdots N-1$ being a unimodular sequence. If $\mu(\mathbf{A}) = \mathcal{O}(1)$, then for all $K$-sparse signals in the time ($\mathbf{\Psi} = \mathbf{I}_N$) or spectral domain ($\mathbf{\Psi} = \frac{1}{\sqrt{N}}\mathbf{F}^*$), $M \geq \mathcal{O}(K \log^4 N)$ measurements are required for uniform recovery; for any given $K$-sparse signal in the time or spectral domain, $M \geq \mathcal{O}(K \log N)$ measurements are needed using $l_1$-based reconstruction.

*Proof:* If $\mu(\mathbf{A}) = \mathcal{O}(1)$, it can be easily derived from Theorem 1 and Theorem 2 that the above statement holds for time-domain $K$-sparse signals (i.e., $\mathbf{\Psi} = \mathbf{I}_N$). To see this is the case in frequency domain (i.e., $\mathbf{\Psi} = \frac{1}{\sqrt{N}}\mathbf{F}^*$), let us examine the coherence parameter $\mu(\mathbf{A\Psi})$. Note that

$$\mathbf{A\Psi} = \frac{1}{\sqrt{N}}\mathbf{F}^*\mathbf{\Sigma}\mathbf{F}\frac{1}{\sqrt{N}}\mathbf{F}^* = \mathbf{F}^*\mathbf{\Sigma}.$$

Obviously the square matrix $\mathbf{F}^*\mathbf{\Sigma}$ is unitary, and all the entries are unimodular, which implies an ideal coherence parameter $\mu\left(\frac{1}{\sqrt{N}}\mathbf{AF}^*\right) = 1$. Therefore, Theorem 3 also holds for spectrally sparse signals. ∎

It is clear that $\mu(\mathbf{A}) = \max(|a_0|, |a_1|, \cdots, |a_N|)$. By (12), the problem now boils down to the construction of a unimodular sequence $\sigma$ so that its normalized IFFT coefficients are bounded by $\mathcal{O}(1)$.

### B. Main Results

In this paper, we study the construction of $\sigma$ from a sequence with good autocorrelation property [25], [26]. Several metrics can be defined to measure the goodness of such sequences, such as the peak sidelobe level, integrated sidelobe level and the merit factor etc. All these metrics can be used for both periodic and aperiodic autocorrelations.

*Definition 2 (Periodic and aperiodic autocorrelations):* For a sequence **s** of period $N$, its periodic autocorrelation $R_\mathbf{s}(l)$ and aperiodic autocorrelation $r_\mathbf{s}(l)$ are respectively defined by

$$R_\mathbf{s}(l) = \sum_{k=0}^{N-1} s_k \cdot s^*_{\mathrm{mod}\ (k+l,N)} \tag{15}$$

$$r_\mathbf{s}(l) = \sum_{k=0}^{N-l-1} s_k \cdot s^*_{k+l} \tag{16}$$

where $l = 0, 1, 2, \cdots$ is an integer.

Let us first consider sequences with small off-peak periodic autocorrelations (or small peak sidelobe level).

*Definition 3 (Perfect and nearly perfect sequences):* A sequence **s** is called a perfect sequence if its periodic autocorrelation $R_\mathbf{s}(l)$ satisfies

$$R_\mathbf{s}(l) = \begin{cases} N & l = iN \\ 0 & l \neq iN. \end{cases} \tag{17}$$

A nearly perfect sequence is a sequence with the off-peak autocorrelation magnitude bounded by a small value $\epsilon$, i.e.,

$$|R_\mathbf{s}(l)| < \epsilon, \quad l \neq iN. \tag{18}$$

Due to their wide applications, the construction of (nearly) perfect sequences has been extensively studied [25], [26]. It is well known that perfect polyphase sequences exist for arbitrary $N$ [27]. But the only known bipolar perfect sequence (i.e., $s_k \in \{1, -1\}$) is $[1, 1, -1, 1]$. For nearly-perfect sequences, a survey on bipolar and quadriphase sequences is given in [28]. A widely used bipolar sequence with $\epsilon = 1$ is the maximum-length sequence (also called as the $m$-sequence), which can be easily implemented by shift registers. Other examples of bipolar sequences with $\epsilon = 1$ are the Legendre sequences and twin-prime sequences. An example of quadriphase sequences with $\epsilon = 1$ is the complementary-based sequence. Sequences with other values of $\epsilon$ such as 2, 3, and 4, can be found in [28]. The following Lemma shows that for a nearly perfect sequence, $\mu(\mathbf{A})$ is bounded by $\sqrt{1+\epsilon}$.

*Lemma 1 (Bound on the coherence parameter):* Let the complex-valued matrix **A** be defined by (11) where $\sigma = \mathbf{s}$. If **s** is a unimodular perfect sequence, then $\mu(\mathbf{A}) = 1$. If **s** is a unimodular nearly perfect sequence satisfying (18), then

$$\mu(\mathbf{A}) \leq \sqrt{1+\epsilon}. \tag{19}$$

*Proof:* First, we examine the FFT $\hat{\mathbf{s}} = \mathbf{Fs}$ of sequence **s**. By the Wiener-Khinchin theorem, the power spectrum $|\hat{\mathbf{s}}|^2$ is given by the FFT of the periodic autocorrelation function $R_\mathbf{s}$. Thus, we have (for $0 \leq k \leq N-1$)

$$\begin{aligned} |\hat{s}_k|^2 &= \sum_{l=0}^{N-1} R_\mathbf{s}(l) e^{-\frac{j2\pi kl}{N}} \\ &\leq N + \left| \sum_{l=1}^{N-1} R_\mathbf{s}(l) e^{-\frac{j2\pi kl}{N}} \right| \\ &\leq N + (N-1)\epsilon. \end{aligned} \tag{20}$$

Now consider the sequence $\mathbf{a} = \frac{1}{\sqrt{N}}\mathbf{F}^*\mathbf{s}$. It is easy to show that $\mathbf{F}^*\mathbf{s}$ is a reversed version of $\hat{\mathbf{s}}$ (with respect to index $k$), hence the same magnitudes. From (20), the coherence parameter, i.e., the peak magnitude of **a**, is bounded by

$$\mu(\mathbf{A}) \leq \frac{1}{\sqrt{N}}\sqrt{N+(N-1)\epsilon} \leq \sqrt{1+\epsilon}.$$

When $\epsilon = 0$, it reduces to the ideal bound $\mu(\mathbf{A}) = 1$ for perfect sequences. ■

Lemma 1 forms the motivation of the method to be developed in this paper. It shows that $\mu(\mathbf{A})$ will be bounded by $\mathcal{O}(1)$ if $\sigma$ is a (nearly) perfect sequence. Note that Lemma 1 is based on periodic autocorrelation. One may wonder whether we have similar results for sequences with good aperiodic autocorrelation? It can be shown that $R_{\mathbf{s}}(l) = r_{\mathbf{s}}(l) + r_{\mathbf{s}}(N-l)$. Thus, if a sequence has bounded $r_{\mathbf{s}}(l)$, $R_{\mathbf{s}}(l)$ is also bounded. However, the design of $\mathbf{s}$ with bounded $r_{\mathbf{s}}(l)$ is much more difficult. Actually, it is impossible to construct sequences with exact impulsive aperiodic autocorrelation (i.e., $r_{\mathbf{s}}(l) = 0$ for $1 \leq l \leq N-1$). Barker sequence has $|r_{\mathbf{s}}(l)| \leq 1$, but it only exists for $N \leq 13$. Several variations have been considered in literature to construct sequences with good aperiodic autocorrelation, such as zero-correlation zone sequences and complementary sequences [28]. In Section IV-C, we will show that $\sigma$ can also be constructed from a complementary sequence.

There are some other limitations of Lemma 1. Firstly, it is difficult, if not impossible to extend to real sensing matrices; secondly, the bound in (19) is pessimistic; thirdly, an extension to other domains (e.g., DCT) seems difficult. All these issues will be addressed in the next section. In a nutshell, Table I lists the unimodular sequences $\sigma$ used in this paper, along with the corresponding $N$ and $\mu(\mathbf{A})$. Among them, the FZC is a perfect polyphase sequence, the $m$-sequence is a bipolar sequence with $\epsilon = 1$ and the Golay sequence is a complementary sequence with good aperiodic autocorrelations. As can be seen, for the $m$-sequence, the bound in Table I is better than $\sqrt{2}$ given by Lemma 1. Table I also suggests that real-coefficient $\mathbf{A}$ with $\mu(\mathbf{A}) = \mathcal{O}(1)$ can be constructed from the extended polyphase or the extended Golay sequences. In addition to these results, we will show in Theorem 4 that if $\sigma$ is chosen as the FZC sequence, similar bounds hold for sparse signals in the DCT domain. Moreover, Theorem 5 shows that if $\mathbf{A}$ is constructed from the $m$-sequence using time-domain approach, it can be used to sample a zero-mean sparse signal.

### C. Potential Applications

Note that sequences with good autocorrelations have already found wide applications that naturally involve convolution. For example, perfect polyphase sequences have been used in Radar pulse compression. The maximum-length and complementary sequences have been used for channel estimation in communications and impulse response (e.g., acoustic or ultrasound system) measurement. The fact that these deterministic sequences can be efficiently used in compressive acquisition of sparse signals allows for new design considerations. It also provides the potential of practical implementation of the CS technology with minimum change in hardware or software. Here, we discuss two examples of potential applications:

Fig. 1(a) illustrates the implementation diagram of the proposed scheme in channel estimation. Here, we assume that the channel response is a time-domain sparse signal with $K$ propagation paths. In single-channel estimation, a length-$N$ deterministic pilot sequence $\mathbf{a}$ is sent directly to probe the channel. In an OFDM system, the pilot sequence $\sigma$ is transmitted on $N$ subcarriers, which is implemented by an $N$-point IFFT (time domain sequence $\mathbf{a}$ is the output of the IFFT). At the receiver, the signal is randomly subsampled with $M$ samples and the channel is estimated using sparse recovery algorithms. In fact, convolutional-based CS sparse channel estimation have been studied in [4], [29], [30], all of which are based on random sequences. As we will show later on, by using a deterministic sequence, not only can we get better theoretical guarantee, it also leads to simplified radio transmitter design in an OFDM system.

Fig. 1(b) shows another example of the proposed system in Fourier Optics or phase coded aperture. The FFT (using the first lens) of the signal is modulated with a deterministic diagonal sequence $\sigma$ (implemented via a spatial light modulator), fed to IFFT (up to a scaling factor $N^{-1/2}$) by the second lens, then randomly subsampled. It is similar to the imaging architecture in [8] but with fixed coefficients of $\sigma$. Simulation results show that our proposed scheme can effectively reconstruct natural images. On the other hand, if we use a random sequence $\sigma$ and sampled deterministically, one can only recover a spatially sparse signal.

Other potential applications of the proposed system include Radar imaging, compressive spectrum measurement and magnetic resonant imaging etc.

### D. Connections With Existing Work

The comparison between the proposed scheme and existing random convolution-based operators is shown in Table II. It can be seen that the proposed scheme offers near optimal theoretical performance guarantee for both uniform and non-uniform CS recovery, thanks to its deterministic construction of $\sigma$. Recall that for a random filter, the coherence parameter is bounded by $\mathcal{O}(\sqrt{\log N})$ [8], [9]. The $\mathcal{O}(1)$ coherence parameter associated with deterministic $\sigma$ enables us to remove the extra $(\log N)$ factor in existing random convolution. Although the proposed scheme can not offer universality, it works for both time and frequency domains (and the DCT domain for FZC sequences). Thus, the proposed scheme can be used as a hardware-friendly, memory-efficient and fast computable solution for large scale CS applications, e.g., hyperspectral imaging. During the revision of this paper, we learnt of the work [23] which presented the same bounds for the partial circulant operator; however, the bounds in [23] hold in the time domain only.

It should be mentioned that deterministic sequences have been investigated in CS before. For example, chirp sequences were applied to radio interferometry in [31], where the sensing matrix was constructed in a different way, namely, it was the product of a rectangular binary matrix $\mathbf{M}$, Fourier matrix $\mathbf{F}$, diagonal matrix $\mathbf{C}$ implementing chirp modulation and diagonal matrix $\mathbf{D}$ implementing the primary beam. The coherence was analyzed when $\mathbf{\Psi}$ is formed by Gaussian waveforms. Chirp sequences were also used to construct deterministic sensing matrices in [32], which cannot be implemented through convolution. Besides, the sizes of the sampling operators in [32] are restricted to be $M \times M^2$. Upon completion of this work, we learned that perfect sequences (including the FZC sequence and real-valued perfect tenary sequences) were used as the entries of Toeplitz sensing matrices in radio spectrum estimation [33]. Yet the analysis of the coherence parameter or RIP was limited to the cases $\boldsymbol{\Psi} = \mathbf{I}_N$ and $\boldsymbol{\Psi} = \frac{1}{\sqrt{N}}\mathbf{F}$ in [33]. Taking a step forward, our results hold for both nearly perfect sequences and complementary sequences. For the particular case of the FZC sequence, we have also generalized the result for the DCT basis. Besides, we have also presented the design of real-valued sensing matrices in Section V with arbitrary length $N$.

TABLE I
COHERENCE PARAMETER $\mu(\mathbf{A})$ FOR DIFFERENT DIAGONAL SEQUENCES $\sigma$

|  | $\sigma$ | $N$ | $\mu(\mathbf{A})$ |
|---|---|---|---|
| Complex matrices | FZC | Arbitrary | 1 |
|  | $m$-sequence | $2^k - 1, k \in \mathbb{N}$ | $\sqrt{1 + \frac{1}{N}}$ |
|  | Golay sequence | $2^{\kappa_1} 10^{\kappa_2} 26^{\kappa_3}, \kappa_1, \kappa_2, \kappa_3 \in \mathbb{N}$ | $\sqrt{2}$ |
| Real matrices | Extended polyphase | Even $N$ | $4 + \frac{4}{\sqrt{N}}$ |
|  |  | Odd $N$ | $2.69 + \frac{8.15}{\sqrt{N}}$ |
|  | Extended Golay | Even $N$, $N = 2^{\kappa_1} 10^{\kappa_2} 26^{\kappa_3}, \kappa_1, \kappa_2, \kappa_3 \in \mathbb{N}$ | $2 + \frac{2}{\sqrt{N}}$ |
|  |  | Odd $N$, $N = 2^{\kappa_1} 10^{\kappa_2} 26^{\kappa_3} \pm 1, \kappa_1, \kappa_2, \kappa_3 \in \mathbb{N}$ | $2 + \frac{1}{\sqrt{N}}$ |

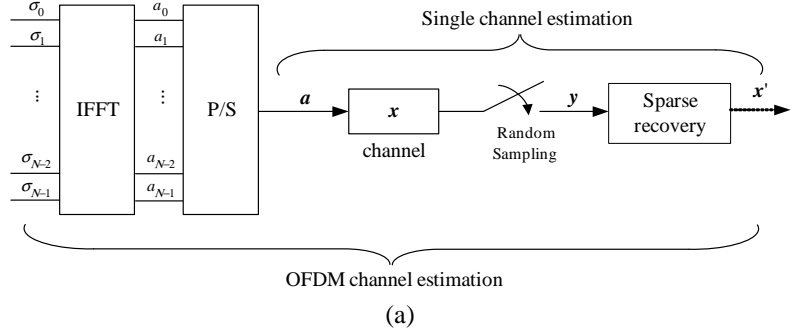

(a)

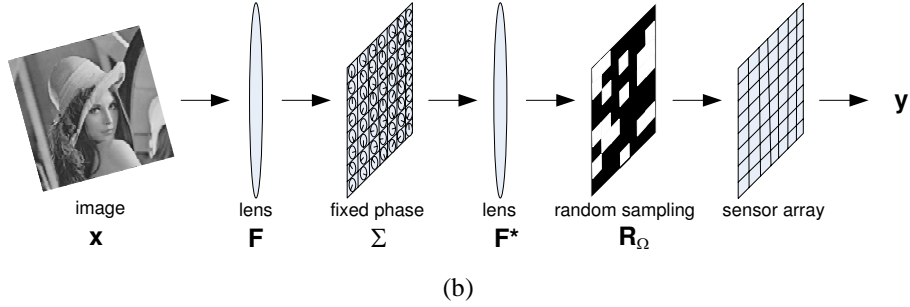

(b)

Fig. 1. Applications of convolutional compressed sensing. (a) Channel estimation; (b) Coded aperture imaging or Fourier optics. The lenses transform the signal to Fourier domain and back.

TABLE II
COMPARISON OF DIFFERENT CIRCULANT OPERATORS FOR COMPRESSED SENSING

| Measurement Operator $\boldsymbol{\Phi}$ | Random Convolution [8] | Partial Circulant Operator [9], [23] | This Work |
|---|---|---|---|
| Filter Coefficients | Random | Random | Deterministic |
| Sub-sampling Operator | Random | Deterministic | Random |
| Sparsifying Transform $\boldsymbol{\Psi}$ | Arbitrary | $\mathbf{I}$ | $\mathbf{I}$ or $\mathbf{IFFT}$ |
| Restricted Isometry Property | $M \geq \mathcal{O}(K \log^5 N)$ | $M \geq \mathcal{O}(K \log^4 N)$ | $M \geq \mathcal{O}(K \log^4 N)$ |
| Non-Uniform Recovery | $M \geq \mathcal{O}(K \log N)$ | $M \geq \mathcal{O}(K \log N)$ | $M \geq \mathcal{O}(K \log N)$ |

## IV. COHERENCE ANALYSIS

The previous section derives the restriction of $\sigma$ under the CS framework. From a practical perspective, there are other restrictions. For example, in Radar, $a_k$ needs to have a constant magnitude. In coded aperture imaging, $a_k$ is real-valued. Also, in OFDM channel estimation, it is desirable that the sequence can minimize the peak power required for signal transmission. It is also preferable that the sequence length is $N = 2^k$ for fast FFT implementation. Some other desirable properties include [34]

- $\sigma$ or $\mathbf{a}$ exists for flexible length $N$;
- $\sigma_k$ or $a_k$ takes a minimum number of values;
- $\sigma_k$ or $a_k$ takes values on a pre-determined signalling set (e.g., BPSK or QPSK);

- $\sigma$ or **a** can be easily generated.

However, none of the existing sequences can satisfy all above mentioned requirements. For instance, the widely used bipolar $m$-sequence has low implementation complexity and good autocorrelation property. However, it only exists for $N = 2^k - 1$ ($k \in \mathbb{N}$). Perfect polyphase (chirp) sequences exist for arbitrary length $N$, but they are difficult to generate. Besides, they are not on the popular constellation sets such as BPSK or QPSK. Taking these facts into account, we carry out a case-by-case coherence analysis of circulant matrices constructed from three classes of sequences: (*i*) the polyphase sequence, (*ii*) $m$-sequence and (*iii*) complementary sequence. Our analysis will be conducted for both complex and real-valued matrices **A**.

### A. Polyphase Sequences

As shown in Lemma 1, when $\sigma$ is a unimodular perfect sequence, we have $|a_k| = 1$, i.e., **A** has the ideal coherence parameter $\mu(\mathbf{A}) = 1$. Such a perfect sequence is said to satisfy the constant amplitude and zero autocorrelation (CAZAC) property. This property is very crucial to applications such as Radar. Here, we study a well known CAZAC sequence, the FZC sequence, that has been used in phase coded Radar and 3GPP Long Term Evolution (LTE). The $\gamma$-th sequence ($\gamma$ is an integer that is coprime with $N$) within the FZC family is given by [35][2]

$$\mathbf{s}_k = \begin{cases} e^{-\frac{j\pi\gamma k^2}{N}}, & \text{for even } N; \\ e^{-\frac{j\pi\gamma k(k+1)}{N}}, & \text{for odd } N, \end{cases} \quad (21)$$

for $k = 0, 1, \cdots, N-1$. In fact, we can set either the filter vector **a** or its FFT coefficient $\sigma$ as the FZC sequence. This is because the Fourier transform of an FZC sequence is another FZC sequence [35], implying both the time and frequency-domain approach apply. Under these constructions, **A** is an unitary circulant matrix with ideal coherence parameter $\mu(\mathbf{A}) = 1$. The next Theorem further shows that matrix **A** generated from the FZC sequence can also be used to sample a sparse signals in the DCT domain.

*Theorem 4:* Let IDCT represent the inverse Type-II DCT matrix, and let **A** be a unitary circulant matrix generated from (11), in which $\sigma$ is an FZC sequence with $\gamma = 1$. The matrix

$$\mathbf{U} = \mathbf{A} \cdot \text{IDCT} = N^{-1/2} \mathbf{F}^* \Sigma \mathbf{F} \cdot \text{IDCT} \quad (22)$$

has coherence parameter $\mu(\mathbf{U}) \leq 6\sqrt{2}$.

The proof of this Theorem relies on the results of partial Gauss sum (summarized in Appendix A) and details will be given in Appendix B. It suggests that an FZC-based operator can be applied in compressive imaging as the Type-II DCT is widely used in image/video compression standards to sparsify the signals. It should also be pointed out that although we could only derive the bound for $\gamma = 1$ in Theorem 4, our simulations indicate that the result may hold for arbitrary $\gamma$. Moreover, **A** constructed from the FZC appears to be incoherent with other bases, such as the wavelet and the modified DCT. How to generalize the above result for different

[2]This definition gives a sequence which is the complex conjugate of the standard one [36]. They obviously have the same autocorrelation magnitudes.

$\gamma$ and different $\Psi$ is an interesting problem that is worth further study.

Next, we will investigate how to generate real-coefficient **a** from a polyphase sequence $\sigma$. Here, $\sigma$ needs to satisfy the conjugate symmetry property in (14). We define an polyphase sequence that resembles the FZC sequence. Specifically, when $N$ is an even number, $\sigma$ is given by

$$\sigma_k = \begin{cases} 1 & k = 0 \\ e^{-j\frac{\pi}{N}k^2} & 1 \leq k \leq \frac{N}{2} - 1 \\ 1 & k = \frac{N}{2} \\ e^{j\frac{\pi}{N}k^2} & \frac{N}{2} + 1 \leq k \leq N - 1; \end{cases} \quad (23)$$

when $N$ is an odd number, $\sigma$ can be expressed as

$$\sigma_k = \begin{cases} 1 & k = 0 \\ e^{-j\frac{\pi}{N}k^2} & 1 \leq k \leq \frac{N-1}{2} \\ -e^{j\frac{\pi}{N}k^2} & \frac{N+1}{2} \leq k \leq N - 1. \end{cases} \quad (24)$$

Unlike the FZC sequence, it is not straightforward to derive the close-form expression of $a_k$ when $\sigma$ is given in (23) or (24). Fortunately, by exploiting partial Gauss sum, we can still show that $a_k$ is bounded by $\mathcal{O}(1)$, as stated in the following Lemma:

*Lemma 2:* Let $\sigma$ be defined as (23) or (24), for even and odd $N$, respectively. Then the coherence parameter satisfies

$$\mu(\mathbf{A}) = \begin{cases} 4 + \frac{4}{\sqrt{N}}, & N \text{ even}; \\ 2.69 + \frac{8.15}{\sqrt{N}}, & N \text{ odd}. \end{cases} \quad (25)$$

The proof can be found in Appendix C. Note that the main advantage of using (23) and (24) is that they exist for arbitrary $N$. Besides, quadratic phase modulation has been proved as an effective method to improve spatial resolution in magnetic resonance imaging [37]. The above lemma implies that in these systems, we can further speed up imaging process by exploiting the CS theory.

### B. Maximum-Length Sequences

Although perfect polyphase sequences exist for arbitrary $N$, in some applications, it is more desirable to use bipolar sequences for $\sigma_k$ due to their easy implementation. A popular bipolar sequence is the *m*-sequence whose autocorrelation is given by [38]

$$R_\mathbf{s}(l) = \begin{cases} N, & l \equiv 0 \mod N; \\ -1, & \text{otherwise}. \end{cases} \quad (26)$$

Accordingly, from the Wiener-Khinchin relation we have

$$|a_k| = \begin{cases} 1, & k = 0; \\ \sqrt{N+1}, & \text{otherwise}. \end{cases} \quad (27)$$

Therefore, if $\sigma$ is chosen as the $m$-sequence, **A** is complex valued with coherence parameter

$$\mu_m(\mathbf{A}) = \sqrt{(N+1)/N}. \quad (28)$$

Unlike the FZC sequence, using frequency-domain approach, we cannot easily get a real-coefficient matrix **A** with $\mu(\mathbf{A}) < \mathcal{O}(1)$ by extending the $m$-sequence. Nonetheless, the time-domain approach can be used here. If we set **a** directly

as an $m$-sequence, the following theorem implies that it can be used to sample a *zero-mean* sparse signal:

*Theorem 5:* Consider an $M \times N$ ($N = 2^k - 1$) CS sampling operator given in (9)-(10), in which **a** is an $m$-sequence. Then, for all zero-mean $K$-sparse signals in the time ($\Psi = \mathbf{I}_N$) or spectral domain ($\Psi = \frac{1}{\sqrt{N}}\mathbf{F}_N^*$), $M \geq \mathcal{O}(K \log^4 N)$ measurements are required for uniform recovery; for any given $K$-sparse signal in the time or spectral domain, $M \geq \mathcal{O}(K \log N)$ measurements are needed using $l_1$-based reconstruction.

*Proof:* It is known that a real-coefficient binary perfect sequence $\tilde{\mathbf{a}}$ can be obtained from an $m$-sequence **a** as follows [34]

$$\tilde{\mathbf{a}} = \sqrt{\frac{N}{N+1}}\mathbf{a} + \left(1 - \frac{1}{\sqrt{N+1}}\right)\mathbf{1}_N, \quad (29)$$

in which $\mathbf{1}_N$ represents an $N \times 1$ all-ones vector. As $\tilde{\mathbf{a}}$ is a perfect sequence, using a similar argument to Lemma 1, we know that $\tilde{\sigma} = \frac{1}{\sqrt{N}}\mathbf{F}\tilde{\mathbf{a}}$ is a unimodular sequence. Hence, the corresponding circulant matrix $\tilde{\mathbf{A}} = \sqrt{\frac{N}{N+1}}\mathbf{A} + \left(1 - \frac{1}{\sqrt{N+1}}\right) \cdot$ ones($N,N$) is orthogonal with $\mu(\tilde{\mathbf{A}}) \leq 2$. Thus, according to Theorem 3, $\tilde{\mathbf{A}}$ can be used for $K$-sparse signals in the time or spectral domain, in which $M \geq \mathcal{O}(K \log^4 N)$ and $M \geq \mathcal{O}(K \log N)$ measurements are required for uniform and non-uniform sparse recovery, respectively. Note that for a zero mean signal **x** (i.e., $\sum x_i = 0$) $\tilde{\mathbf{A}}\mathbf{x} = \sqrt{\frac{N}{N+1}}\mathbf{A}\mathbf{x}$. Hence, measurement using **A** is equivalent to that of $\tilde{\mathbf{A}}$ except for a scaling factor $\sqrt{\frac{N}{N+1}}$. As a result, Theorem 5 holds. ∎

Note that zero mean is a very mild restriction. Also, the DC component of a signal can be easily measured, implying the proposed scheme is applicable to nonzero-mean signals.

The $m$-sequence has found applications in measurement of impulse response, channel estimation, spread spectrum communications and fMRI. In the classical method of impulse response measurement, the measured data need to be sampled at a full rate and reconstructed using the correlation method. If the impulse response (or its transfer function) is sparse, the above theorem suggests that it can be reconstructed from a small random subset of measurements. In other words, in case when there is impulsive noise or erasure errors, sparse reconstruction can provide robust recovery. It should be pointed out that due to the close connection between the $m$-sequence and Walsh-Hadamard operator [39], multiplication of **A** can be easily implemented using $\mathcal{O}(N \log N)$ additions.

*Remark 1:* Other (nearly) perfect sequences, for example, the tenary perfect sequences [33], [40] and the Legendre sequence [28] (with $\epsilon = 1$), could also be used to construct an orthogonal **A** using time domain approach. Yet, they are not as widely used as the $m$-sequence. Besides, the lengths of these sequences are not quite flexible. For a tenary perfect sequence, $N = \frac{p^{sk}-1}{p^s-1}$, in which $p$ is an odd prime, $k$ and $s$ are integers. The Legendre sequence only exists for odd prime $N$ [41].

### C. Golay Sequences

Note that the $m$-sequence only exists for $N = 2^k - 1$. As mentioned before, in many applications, such as OFDM, it is desired to have $N = 2^k$. In this subsection, we consider binary complementary sequences (also known as Golay sequences) introduced by Golay in 1961 [42]. These sequences have good aperiodic autocorrelation and they have found numerous applications in spectroscopy, ultrasound and acoustic measurement, Radar pulse compression, WI-FI networks, OFDM and non-destructive test etc. [43], [44].

*Definition 4 (Golay sequences):* Let $\mathbf{a} = (a_0, a_1, \cdots, a_{N-1})$ and $\mathbf{b} = (b_0, b_1, \cdots, b_{N-1})$ be a pair of binary sequences with values 1 or $-1$. Then **a** and **b** form a Golay complementary pair if

$$r_{\mathbf{a}}(l) + r_{\mathbf{b}}(l) = \begin{cases} 2N, & l = 0 \\ 0, & l = 1, \cdots, N-1. \end{cases} \quad (30)$$

A sequence in any complementary pair is called a *Golay sequence*.

Let $S_a(\omega) = \sum_{n=0}^{N-1} a_k e^{j\omega n}$ and $S_b(\omega) = \sum_{n=0}^{N-1} b_k e^{j\omega n}$ represent the discrete time Fourier transform of **a** and **b**, respectively. Then, the following relation holds

$$|S_{\mathbf{a}}(\omega)|^2 + |S_{\mathbf{b}}(\omega)|^2 = 2N. \quad (31)$$

Note that for any length-$N$ Golay sequence **s**, it simply follows from (31) that $|S_{\mathbf{s}}(\omega)|^2 \leq 2N$. Based on this property, we can easily arrive at the following corollary:

*Corollary 1:* Let the diagonal sequence $\sigma$ be a Golay sequence **s** of length $N$. Then, **A** is a unitary circulant matrix with $\mu(\mathbf{A}) \leq \sqrt{2}$.

As mentioned previously, one promising application of convolutional CS is in OFDM channel estimation. One of the main practical issues of the OFDM is the Peak-to-Average Power Ratio (PAPR) of the transmitted signal, which can significantly impact the efficiency of the power amplifier. Specifically, for a length-$N$ sequence $\sigma$, its PAPR is given by [45]

$$\text{PAPR}(\sigma) = \max_{t \in [0,T)} \left| \frac{1}{\sqrt{N}} \sum_{n=0}^{N-1} \sigma_n e^{\frac{j2\pi nt}{T}} \right|^2,$$

where $T$ is the OFDM symbol period. For a Golay sequence, it is obvious that its PAPR is bounded by 2. In contrast, if $\sigma$ is a random sequence, its PAPR is asymptotically $\log N$ with probability 1 [45]. From this point of view, Golay sequences are more advantageous for compressive OFDM channel estimation than random sequences.

We then move on to consider real sensing matrices constructed from an extended Golay sequence.

*Theorem 6 ($\mu(\mathbf{A})$ for extended Golay sequence):* Let **s** be a Golay sequence with length of $N_0$. When $N = 2N_0$, $\sigma$ is define by

$$\sigma_k = \begin{cases} s_k & 0 \leq k \leq \frac{N}{2} - 1, \\ s_0 & k = \frac{N}{2}, \\ s_{N-k} & \frac{N}{2} + 1 \leq k \leq N - 1; \end{cases} \quad (32)$$

When $N = 2N_0 + 1$, $\sigma$ is defined by

$$\sigma_k = \begin{cases} s_k & 0 \leq k \leq \frac{N-1}{2} - 1, \\ s_{N-k} & \frac{N+1}{2} \leq k \leq N - 1; \end{cases} \quad (33)$$

Then $\mu(\mathbf{A})$ is bounded by

$$\mu(\mathbf{A}) \leq \begin{cases} 2 + \frac{2}{\sqrt{N}}, & N \text{ even}; \\ 2 + \frac{1}{\sqrt{N}}, & N \text{ odd}. \end{cases} \quad (34)$$

*Proof:* We show the proof for even $N$, while the case of odd $N$ is omitted since it is very similar. Note that

$$\begin{aligned}\mu(\mathbf{A}) &= \frac{1}{\sqrt{N}} \max_{0 \leq k < N} \left| \sum_{i=0}^{N-1} \sigma_i e^{\frac{j2\pi}{N} ki} \right| \\ &= \frac{1}{\sqrt{N}} \max_{0 \leq k < N} \left| S_{\mathbf{s}}\left(\frac{2\pi k}{N}\right) + S_{\mathbf{s}}\left(-\frac{2\pi k}{N}\right) + s_0 e^{j\pi k} - s_0 \right| \\ &\leq \frac{2}{\sqrt{N}} \left[ \max_{0 \leq k < N} \left| S_{\mathbf{s}}\left(\frac{2\pi k}{N}\right) \right| + 1 \right] \end{aligned} \quad (35)$$

in which $S_{\mathbf{s}}(\omega)$ is the discrete-time Fourier transform of $\mathbf{s}$, i.e., $S_{\mathbf{s}}(\omega) = \sum_{k=0}^{N_0-1} s_k e^{j\omega k}$. By definition of the Golay complementary sequence, $|S_{\mathbf{s}}(\omega)| \leq \sqrt{2N_0} = \sqrt{N}$. Consequently, $\mu(\mathbf{A}) \leq 2 + \frac{2}{\sqrt{N}}$ for $N = 2N_0$. ∎

The above result indicates that the extended Golay sequence is quite attractive in phase coded aperture imaging. As $\sigma_k$ is bipolar, it can be easily implemented using a spatial light modulator. Besides, Golay sequences are known to exist for all lengths $2^{\kappa_1} 10^{\kappa_2} 26^{\kappa_3}$, $\kappa_1, \kappa_2, \kappa_3$, which offers great flexibility in image sizes.

*Remark 2 (Golay-OSTM):* The authors introduced orthogonal symmetric Toeplitz matrices (OSTM) as sensing matrices in [46], and proposed to use the Golay sequence as the diagonal sequence in [47]. In general, OSTM may be viewed as a special case of convolutional CS, where the sequence $\mathbf{s} \in \{-1, 1\}^{N/2} \setminus \{(-1, -1, \cdots, -1), (1, 1, \cdots, 1)\}$. The symmetry of OSTM, i.e., $\mathbf{A}(i,j) = \mathbf{A}(j,i) = a_{|i-j|}$, could can be exploited for fast reconstruction.

## V. SIMULATION RESULTS

In this section, we present the simulation results for the following two experiments:

*Experiment 1*: In this experiment, we aim to study the potential of the proposed system in OFDM channel estimation. Here, the number of carrier is $N = 1024$ and $M = 64$ samples are obtained at the receiver. $\sigma$ is set to be the Golay sequence due to its excellent PAPR. The channel model is the ATTC (Advanced Television Technology Center) and the Grande Alliance DTV laboratorys ensemble E model, whose static case impulse response $\mathbf{x}(n)$ can be expressed as [48]

$$\begin{aligned}\mathbf{x}(n) =& \delta(n) + 0.3162\delta(n-2) + 0.1995\delta(n-17) \\ &+ 0.1296\delta(n-36) + 0.1\delta(n-75) + 0.1\delta(n-137);\end{aligned} \quad (36)$$

We consider the cases when the signal to noise ratios (SNRs) are 0 dB, 10 dB, 20 dB and 30 dB, respectively. For each SNR, 500 trials have run using the subspace pursuit [18] algorithm. The recovery performance is compared with that proposed in [30], in which $\sigma$ is a unimodular sequence with random phase and $\mathbf{R}_\Omega$ is a deterministic sampling operator. From Table III, it can be seen that both schemes offer similar reconstruction

TABLE IV
CIRCULANT MATRICES USED IN EXPERIMENT 2

| Schemes | $\sigma$ | $\mathbf{R}_\Omega$ |
|---|---|---|
| RP+DS | Random phase | Deterministic |
| RP+RS | Random phase | Random |
| E-Poly+RS | Extended polyphase sequence | Random |
| E-Golay+RS | Extended Golay sequence | Random |

performance. Yet, with the Golay sequence, the PAPR is much lower than that of random sequence.

*Experiment 2*: Here, we study the application of compressive imaging of 2D signals using coded aperture imaging or Fourier optics. In these applications, $a_k$ ($k = 0, \cdots, N-1$) need to be real-coefficient. Hence, the extended polyphase sequence and the extended Golay sequence are considered. They are compared with that of unimodular random phase sequences with random and deterministic sampling, as listed Table IV. The fast reconstruction algorithm for Toeplitz matrices in [11] is applied. Three 8-bit, $256 \times 256$ test images have been used, as shown in Fig. 2. In particular, Fig. 2(a) "Star-sky" is a spatially sparse astronomical image, Fig. 2(f) "Tempel Comet" is a smooth astronomical image, and Fig. 2(k) "Brain" is a medical image. The recovered images at $R = 0.25$ bits per pixel (bpp) are shown in Fig. 2 and the reconstructed SNRs at different bit rates are presented in Fig. 3. As can be seen, when deterministic sampling is used, the reconstructed SNRs are much lower than those using random sampling. For a spatially sparse image "Star Sky", the visual difference is small. But for the other two spectrally sparse images, the scheme using random sequence with deterministic sampling failed to recover the original image. These results suggested that partial random circulant matrices with deterministic sampling are not efficient in acquisition of natural images. In contrast, the proposed schemes with deterministic sequences can offer similar performance as those of random sequences with random sampling, both in terms of the SNRs and the visual quality of reconstructed images.

## VI. CONCLUSIONS

In this paper, we have proposed a new class of circulant sensing matrices, which were constructed from deterministic sequences with good autocorrelation property, such as the FZC sequence and Golay sequence. We showed that these convolutional sensing matrices have a small coherence parameter with the time and spectral domain, so that the original sparse signal could be faithfully recovered. The proved theoretic guarantee is the strongest among existing convolution-based CS schemes. Experimental results show that these sensing matrices compare favorably with existing structured random matrices. Since the underlying deterministic sequences are widely used in practice, a major advantage of the proposed scheme is that it may be integrated into existing systems. Namely, when part of the data is corrupted such that traditional techniques fail to recover the original signal, our analysis shows that the CS recovery algorithms will be helpful. It is hoped that this work will open the door to more applications of sequences in CS.

TABLE III
AVERAGE OUTPUT SNRs IN OFDM CHANNEL ESTIMATION

|  |  | Partial Circulant Operator | Proposed System |
|---|---|---|---|
| $\sigma$ | | Random Phase | Golay |
| $\mathbf{R}_\Omega$ | | Deterministic | Random Sampling |
| PAPR | | [6.4, 15.6] | 2 |
| Input SNR | 0 dB | 5.32 dB | 5.44 dB |
|  | 10 dB | 13.98 dB | 14.34 dB |
|  | 20 dB | 37.53 dB | 37.48 dB |
|  | 30 dB | 45.22 dB | 45.61 dB |

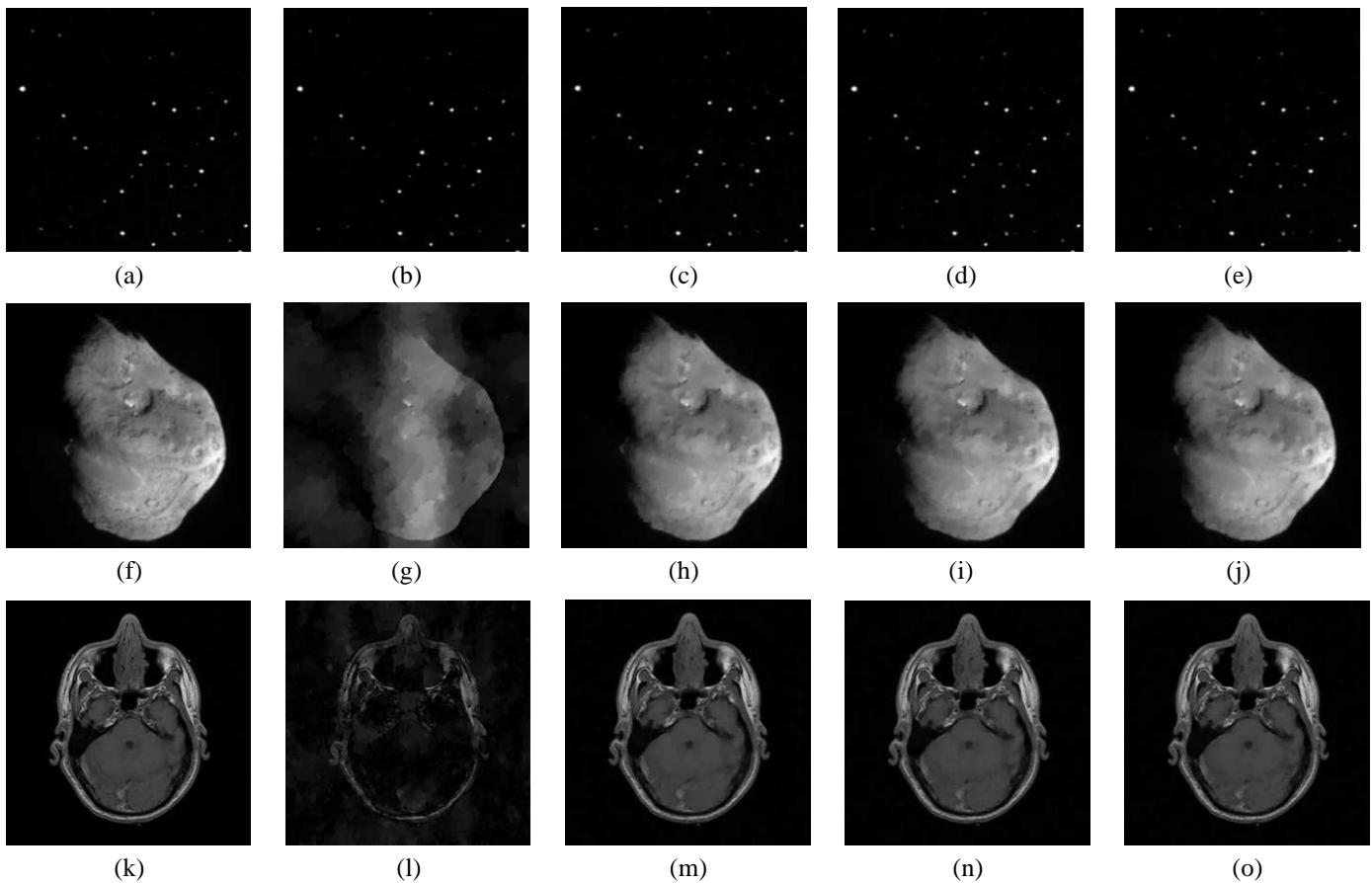

Fig. 2. Original $256 \times 256$ images and reconstructed results at 0.25bpp using different real-coefficient images. First row, "Star-sky" images. (a) Original image; (b) RP+DS: 17.85 dB; (c) RP+RS: 24.02 dB; (d) E-Poly+RS: 24.10 dB; (e) E-Golay+RS: 24.11 dB. Second row, "Tempel Comet" images. (f) Original image; (g) RP+DS: 0.37 dB; (h) RP+RS: 30.16 dB; (i) E-Poly+RS: 29.88 dB; (j) E-Golay+RS: 30.26 dB. Third row, "Brain" images. (k) Original image; (l) RP+DS: 1.00 dB; (m) RP+RS: 20.93 dB; (n) E-Poly+RS: 20.94 dB; (o) E-Poly+RS: 20.98 dB. Please refer to Table IV for acronyms of different schemes.

## APPENDIX A
## PARTIAL GAUSS SUMS

*Definition 5:* Let $N$ be a positive integer. The exponential sum [49]

$$G_N(m) = \sum_{k=0}^{m-1} e^{j2\pi k^2/N} \quad (37)$$

is an *incomplete Gauss sum* when $m < N$.

When $m = N$, the complete Gauss sum $G_N(N)$ is well known [49]

$$G_N(N) = \begin{cases} (1+j)\sqrt{N}, & \text{if } N \equiv 0 \pmod{4} \\ \sqrt{N}, & \text{if } N \equiv 1 \pmod{4} \\ 0, & \text{if } N \equiv 2 \pmod{4} \\ j\sqrt{N}, & \text{if } N \equiv 3 \pmod{4}. \end{cases} \quad (38)$$

Moreover, when $m \leq (N+1)/2$,

$$G_N(m) + G_N(N-m+1) = 1 + G_N(N). \quad (39)$$

Define a normalized version $g_N(m)$ as

$$g_N(m) = 2N^{-\frac{1}{2}} \sum_{k=0}^{m} e^{j2\pi k^2/N}, \quad (40)$$

then we have [49]

$$|g_N(m)| \leq \begin{cases} \sqrt{2}, & \text{if } N = 4k, m \leq N/2 \\ 1.07 + \mathcal{O}(N^{-\frac{1}{2}}), & \text{if } N = 4k+1, m < N/2 \\ 0.95 + \frac{101}{40}N^{-\frac{1}{2}}, & \text{if } N = 4k+2, m \leq N \\ \sqrt{1+N^{-1}}, & \text{if } N = 4k+3, m < N/2, \end{cases} \quad (41)$$

where $k$ is a positive integer.

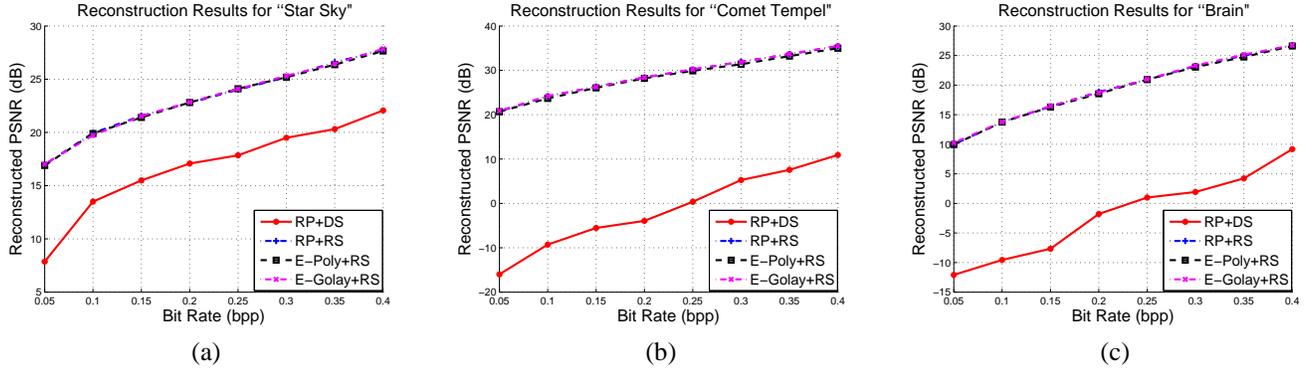

Fig. 3. SNR reconstruction results for three $256 \times 256$ images using different real-coefficient circulant matrices. (a) "Star sky" image; (b) "Comet Tempel" image; (c) "Brain" image.

Based on the above results, we can easily arrive at the following Lemma:

*Lemma 3:* Define $G_{2N}(m)$ as

$$G_{2N}(m) = \sum_{k=0}^{m-1} e^{j\frac{2\pi}{2N}k^2} = \sum_{k=0}^{m-1} e^{j\frac{\pi}{N}k^2}.$$

Then, $G_{2N}(m)$ can be bounded

$$|G_{2N}(m)| \leq \begin{cases} \sqrt{N}, & \text{if } N = 2N_0, m = 1, N \\ 3\sqrt{N}+1, & \text{if } N = 2N_0, N+1 \\ \frac{\sqrt{2N}}{2} \cdot \left(0.95 + \frac{101}{40}N^{-\frac{1}{2}}\right), & \text{if } N = 2N_0+1, m \leq 2N, \end{cases} \tag{42}$$

## APPENDIX B
## PROOF OF THEOREM 4

*Proof:* For clarity we only give the proof for even $N$ here. The case of odd $N$ is similar and is omitted. We apply a result from [35] that the Fourier dual of a unimodular perfect sequence yields another unimodular perfect sequence. Specifically, when $\sigma$ is an FZC sequence in (21) with $\gamma = 1$, the elements of $\mathbf{A}$ are given by [35]

$$\mathbf{A}(p,q) = e^{\frac{j\pi(p-q)^2}{N} - \frac{j\pi}{4}}. \tag{43}$$

Obviously, we may ignore the phase $e^{-\frac{j\pi}{4}}$ when calculating its magnitude. Note that the IDCT coefficients are can be expressed as

$$\text{IDCT}(p,q) = \begin{cases} \frac{1}{\sqrt{N}}, & q = 0 \\ \sqrt{\frac{2}{N}} \cos\left(\frac{\pi}{N}(p+\frac{1}{2})q\right), & 1 \leq q \leq N-1. \end{cases} \tag{44}$$

Thus, when $q = 0$ and $0 \leq p \leq N-1$,

$$|\mathbf{U}(p,0)| = \frac{1}{\sqrt{N}} \left| \sum_{k=0}^{N-1} e^{j\frac{\pi}{N}(p-k)^2 - \frac{j\pi}{4}} \right| \tag{45}$$
$$= \frac{1}{\sqrt{N}} |G_{2N}(N)| \leq 1.$$

The last step is due to (42). When $1 \leq q \leq N-1$ and $0 \leq p \leq N-1$,

$$\mathbf{U}(p,q) = \sum_{k=0}^{N-1} \mathbf{A}(p,k) \cos\left(\frac{\pi}{N}(k+\frac{1}{2})q\right) \cdot \frac{\sqrt{2}}{\sqrt{N}}$$
$$= \frac{\sqrt{2}}{2\sqrt{N}} \sum_{k=0}^{N-1} e^{j\frac{\pi}{N}(p-k)^2 - \frac{j\pi}{4}} \left(e^{-j\frac{\pi}{N}(k+\frac{1}{2})q} + e^{j\frac{\pi}{N}(k+\frac{1}{2})q}\right). \tag{46}$$

Through some mathematical manipulations, it can be shown that $\mathbf{U}(p,q)$ can be bounded by

$$|\mathbf{U}(p,q)| \leq \frac{\sqrt{2}}{2\sqrt{N}} \left( \left| \sum_{k=0}^{N-1} e^{j\frac{\pi}{N}(k-p-\frac{1}{2}q)^2} \right| + \left| \sum_{k=0}^{N-1} e^{j\frac{\pi}{N}(k-p+\frac{1}{2}q)^2} \right| \right)$$
$$\leq \sqrt{\frac{2}{N}} \left| \sum_{k=0}^{N-1} e^{j\frac{\pi}{N}(k-p-\frac{1}{2}q)^2} \right| \tag{47}$$

When $q = 2q_0$ is an even number, let $p + q_0 = l, 0 \leq l \leq \frac{3N}{2} - 1$, we have $e^{j\frac{\pi}{N}(k-p-\frac{1}{2}q)^2} = e^{j\frac{\pi}{N}(k-l)^2}$. If $0 \leq l \leq N-1$,

$$\left| \sum_{k=0}^{N-1} e^{j\frac{\pi}{N}(k-l)^2} \right| = |G_{2N}(l+1) + G_{2N}(N-l) - 1| \leq 2\sqrt{N}+1 \tag{48}$$

where the last step is due to partial Gaussian Sum. If $N \leq l \leq \frac{3N}{2} - 1$,

$$\left| \sum_{k=0}^{N-1} e^{j\frac{\pi}{N}(k-l)^2} \right| \leq |G_{2N}(2N-l) + G_{2N}(l-N+1) - 1| \leq 2\sqrt{N}+1 \tag{49}$$

where again the property $|G_{2N}(l)| \leq \sqrt{N}$ for $l \leq N$ is applied. So for even $q$, we have $|\mathbf{U}(p,q)| \leq 2\sqrt{2} + \frac{\sqrt{2}}{\sqrt{N}}$. When $q = 2q_0 + 1$ is an odd number, let $l = p + q_0$. After some tedious calculations, we may break up (47) into

$$|\mathbf{U}(p,q)| \leq \frac{\sqrt{2}}{2\sqrt{N}} (|Q_N(\text{mod}(l,N)+1)| + |Q_N(N-1-\text{mod}(l,N))| + |Q_N(|p-q_0|)| + |Q_N(N-|p-q_0|)|), \tag{50}$$

where $Q_N(m)$ is given by

$$Q_N(m) = \sum_{k=0}^{m-1} e^{j\frac{\pi}{N}(k+\frac{1}{2})^2} = \sum_{k=0}^{m-1} e^{j\frac{\pi}{4N}(2k+1)^2}, 0 \leq m \leq N \tag{51}$$

It can be shown that $Q_N(m) = G_{8N}(2m) - G_{2N}(m)$, which leads to

$$|Q_N(m)| \leq |G_{8N}(2m)| + |G_{2N}(m)| = 3\sqrt{N} \tag{52}$$

for $0 \leq m \leq N$. As a result, $|\mathbf{U}(p,q)| \leq 6\sqrt{2}$ for odd $q$. Combined with the result for even $q$, we arrive at Theorem 4. ∎

## APPENDIX C
## PROOF OF LEMMA 2

*Proof:* The proof is also based on the incomplete Gauss sum in Appendix A. Again, for simplicity, we only present the proof for even $N$.

As $\sigma_{N-k} = \sigma_k^*, 1 \leq k \leq \frac{N}{2} - 1$, it can be shown that $a_l$ takes the following form

$$a_l = \frac{1}{\sqrt{N}} \sum_{k=0}^{N/2-1} e^{\frac{j\pi}{N}(-k^2+2lk-l^2)} \cdot e^{\frac{j\pi}{N}l^2}$$
$$+ \frac{1}{\sqrt{N}} \sum_{k=0}^{N/2-1} e^{\frac{j\pi}{N}(k^2-2lk+l^2)} \cdot e^{-\frac{j\pi}{N}l^2} - \frac{1}{\sqrt{N}} + \frac{(-1)^l}{\sqrt{N}}.$$

Thus, we have

$$|a_l| \leq \frac{2}{\sqrt{N}} \left| \sum_{k=0}^{N/2-1} e^{\frac{j\pi}{N}(k-l)^2} \right| + \frac{2}{\sqrt{N}}, \quad 0 \leq l \leq N-1. \tag{53}$$

When $0 \leq l \leq \frac{N}{2} - 1$,

$$|a_l| \leq = \frac{2}{\sqrt{N}} \left| G_{2N}(l+1) + G_{2N}\left(\frac{N}{2} - l\right) - 1 \right| + \frac{2}{\sqrt{N}}$$
$$\leq \frac{2}{\sqrt{N}} \left( 2\sqrt{N} + 1 \right) + \frac{2}{\sqrt{N}} \leq 4 + \frac{4}{\sqrt{N}}. \tag{54}$$

In a similar way, it can be shown that $|a_{N-l}|$ for $0 \leq l \leq \frac{N}{2} - 1$ can be bounded by

$$|a_{N-l}| \leq \frac{2}{\sqrt{N}} \left( \left| G_{2N}\left(\frac{N}{2} + l\right) \right| + |G_{2N}(l)| \right) + \frac{2}{\sqrt{N}}$$
$$\leq \frac{2}{\sqrt{N}} \cdot 2\sqrt{N} + \frac{2}{\sqrt{N}} = 4 + \frac{2}{\sqrt{N}}, \tag{55}$$

which completes the proof. ∎